\documentclass[modern]{aastex61}

\usepackage{graphicx}
\usepackage{amsmath}
\usepackage{amssymb}
\usepackage{enumitem}

\newcommand\mybar{\kern1pt\rule[-\dp\strutbox]{.8pt}{\baselineskip}\kern1pt}

\setlist[itemize]{noitemsep, topsep=0pt, leftmargin=*}

\shorttitle{Galactic Kites}
\shortauthors{Loeb}

\begin{document}

\title{Galactic Kites}

\author{Abraham Loeb}
\affiliation{Astronomy Department, Harvard University, 60 Garden
  St., Cambridge, MA 02138, USA}

\begin{abstract}
I show that interstellar films of material thinner than a micron,
drift away from the Galactic plane as a result of stellar radiation
pressure.  Such films, whether produced naturally by dust coagulation
in proto-planetary disks or artificially by technological
civilizations, would accumulate over the age of the Milky-Way and
hover above the Galactic disk at a scale-height set gravitationally by
the dark matter halo. Limits on scattered starlight imply that this
population carries a fraction below $2\times 10^{-3}$ of the
interstellar medium mass.

\end{abstract}

\section{Introduction}

The luminosity per unit mass of the Sun is $\sim 3\times 10^{-5}$ of
the Eddington limit, at which the outward radiative push equals the
inward gravitational pull on ionized hydrogen
\citep{1983bhwd.book.....S}. Therefore, an increase by a factor bigger
than $1/(3\times 10^{-5})$ in the cross-sectional area per unit mass
over the fiducial value of $(\sigma_T/m_p)=0.4~{\rm cm^2~g^{-1}}$,
would make the radiative force larger than the gravitational
force. Here, $\sigma_T=0.67\times 10^{-24}~{\rm cm^2}$ is the Thomson
cross-section for electron scattering and $m_p=1.67\times
10^{-24}~{\rm g}$ is the proton mass. Under optically-thin conditions,
both the radiative and gravitational forces scale inversely with the
square of the distance from a star. Therefore, their ratio is
independent of position.

For a thin flat film of solid material with a mass density $\rho_s$
and thickness $w$, the face-on area-per-unit-mass equals
$(1/w\rho_s)$. The required enhancement to a value $\gtrsim
1.2\times 10^4~{\rm cm^2~g^{-1}}$ so that radiation dominates gravity,
is possible at a thickness $w$ satisfying,
\begin{equation}
w<w_c=0.8 \mu{\rm m} \left({\rho_s\over 1~{\rm
    g~cm^{-3}}}\right)^{-1} .
\label{one}
\end{equation}
Instabilities could lead to tangled, non-planar configurations or to
arbitrary orientations of the film \citep{2017ApJ...837L..20M}, which
do not provide the full frontal area of the film facing the radiation
source. Below we refer to the ``effective width'', ${\bar w}$,
corresponding to the value that $w$ should have had for face-on
orientation after averaging over all geometries of the population of
films.

For the thin disk of Milky-Way stars, the repulsive radiative force
away from the midplane would exceed the attractive gravitational force
for ${\bar w}< w_c$. The surface mass density of interstellar gas is
smaller than that of stars \citep{2015ApJ...814...13M} and is ignored
in our order-of-magnitude considerations here.

The local surface density of stars, $\Sigma_\star \approx 30~{\rm
  M_\odot~pc^{-2}}$ \citep{2015ApJ...814...13M}, yields the
gravitational acceleration towards the midplane of the stellar disk,
\begin{equation}
g_\star=G\Sigma_\star= 5\times 10^{-10}~{\rm cm~s^{-2}}.
\label{two}
\end{equation}
The mass-to-light ratio of the local disk is of order unity in solar
units \citep{2006MNRAS.372.1149F}, hence maintaining the numerical
coefficients mentioned above for the Sun.

For a face-on film of material, the radiative acceleration away from
the disk midplane exceeds $g_\star$ by a factor $(w_c/{\bar w})$. As a
result of friction with the interstellar medium, the film would
develop a drift speed perpendicular to the disk plane, $v$, at which
the radiative acceleration, $a_{\rm rad}=g_\star(w_c/{\bar w})$, is balanced
by the ram-pressure deceleration induced by the ambient gas. For
${\bar w}\ll w_c$, the two accelerations scale in proportion to the
area-per-unit-mass, yielding
\begin{equation}
a_{\rm rad}= \left({\rho_{\rm ISM}\over \rho_s}\right)\left({v^2 \over
  {\bar w}}\right) ,
\label{three}
\end{equation}
where $\rho_{\rm ISM}\approx m_p n_{\rm ISM}$ is the mass density of the
interstellar medium; for a fiducial proton number density of $n_{\rm
  ISM}\sim 1~{\rm cm^{-3}}$ we get $\rho_{\rm ISM}\sim
10^{-24}\rho_s$ \citep{2011piim.book.....D}.

\section{Results}

Equations (\ref{one})-(\ref{three}) imply a drift speed of $v\sim
2~{\rm km~s^{-1}}(n_{\rm ISM}/1~{\rm cm^{-3}})^{-1/2}$, away from the
Galactic midplane.  At this subsonic drift speed, ambient gas
particles deposit an energy equivalent of $\sim 2\times 10^{-2}~{\rm
  eV}=260~{\rm K}$ per proton, making such impacts insignificant
relative to thermal effects \citep{2018ApJ...860...42H}.

At a sufficiently high elevation relative to the midplane, the
gravitational force from the dark matter halo binds the
film population. Since the local mass density of dark matter is
comparable to that of stars \citep{2019JCAP...04..026B}, the
scale-height of the film population $h_{\rm f}$ would exceed that of stars
$h_\star$ (of order a few hundred pc) by a factor of $(a_{\rm
  rad}/g)\sim (w_c/{\bar w})$.

The time it takes films to populate this scale-height is of order
$(h_{\rm f}/v)\sim 3\times 10^8~{\rm yr}$, much shorter than the age
of the Galaxy.

\section{Implications}

Thin films of the required surface-area per unit mass could be
produced naturally by coagulation of dust particles in the midplane of
protoplanetary disks \citep{2019ApJ...872L..32M,2020ApJ...900L..22L}
or artificially by technological civilizations like ours
\citep{2018ApJ...868L...1B,2021arXiv211015213L}.

The potential population of films above the Galactic disk would
scatter starlight. Limiting their cumulative optical-depth for
scattering to be below observed limits \citep{2022ApJ...927L...8L}, we
find that the fraction, $f_g$, of the interstellar gas mass incorporated
into thin films of effective width, ${\bar w}$, is limited to the small
value, well below the abundance of heavy elements,
\begin{equation}
f_{\rm g}< 2\times 10^{-3}~\left({{\bar w}\over w_c}\right) ,
\label{four}
\end{equation}

Interstellar films with ${\bar w}\gg w_c$ \citep{2018ApJ...868L...1B},
would enter the solar system and could be identified by the upcoming
{\it Legacy Survey of Space and Time} (LSST) on the Vera C. Rubin
Observatory \citep{2022ApJS..258....1B}, as they would reflect
sunlight during their passage close to Earth, similarly to the first
reported interstellar object 1I/2017 U1/`Oumuamua
\citep{2017Natur.552..378M}.

\bigskip
\bigskip
\section*{Acknowledgements}

This work was supported in part by a grant from the Breakthrough Prize
Foundation and by Harvard's {\it Black Hole Initiative} which is
funded by grants from GBMF and JTF. I thank Morgan MacLeod and Sunny
Vagnozzi for insightful comments on the manuscript.

\bigskip
\bigskip
\bigskip

\bibliographystyle{aasjournal}
\bibliography{k}
\label{lastpage}
\end{document}